# Anomalous roughness with system size dependent local roughness exponent

Alexander S. Balankin[1,2,3] and Daniel Morales Matamoros[2,3]

[1)] SEPI, ESIME, Instituto Politécnico Nacional, México D.F. 07738

[2)] Grupo "Mecánica Fractal", México, http://www.mfractal.esimez.ipn.mx

[3)] Instituto Mexicano de Petróleo, México D.F. 07730

We note that in a system far from equilibrium the interface roughening may depend on the system size which plays the role of control parameter. To detect the size effect on the interface roughness, we study the scaling properties of rough interfaces formed in paper combustion experiments. Using paper sheets of different width $\lambda L_0$, we found that the turbulent flame fronts display anomalous multi-scaling characterized by non universal global roughness exponent $\alpha$ and the system size dependent spectrum of local roughness exponents, $\zeta_q(\lambda) = \zeta_1(1) q^{-\omega} \lambda^\phi < \alpha$, whereas the burning fronts possess conventional multi-affine scaling. The structure factor of turbulent flame fronts also exhibit unconventional scaling dependence on $\lambda$. These results are expected to apply to a broad range of far from equilibrium systems, when the kinetic energy fluctuations exceed a certain critical value.

68.35.Fx, 05.40.-a, 05.70.Ln, 61.43.Hv

# 1. INTRODUCTION AND BACKGROUND

Kinetic roughening of interfaces occurs in a wide variety of physical situations,[1] ranging from the fluid invasion in porous media,[2] crystal growth,[3] and motion of flux lines in superconductors,[4] to the meandering fire front propagation in a forest[5] and fracture phenomena.[6] In many cases of interest, a growing interface can be represented by a single-valued function $z(x,t)$ giving the interface location at position $x$ at time $t$ [1-6].

Extensive theoretical and experimental studies of the last decade have led to a classification of interface roughening according to the asymptotic behavior of such quantities as the local and the global interface width [1], defined as

$$w = \left\langle \left\langle \left[ z(x,t) - \langle z(x,t) \rangle_\Delta \right]^2 \right\rangle_\Delta \right\rangle_R^{1/2} \text{ and } W(L,t) = \left\langle \overline{\left[ z(x,t) - \bar{z}(t) \right]^2} \right\rangle_R^{1/2},$$

respectively; here $\langle ... \rangle_\Delta$ is a spatial average over axis $x$ in a window of size $\Delta$, the overbar average denotes over all $x$ in a system of size $L$, and $\langle ... \rangle_R$ denotes average over different realizations. Alternatively, the interface roughness can be characterized by calculating the structure factor or power spectrum, $S(k,t) = \langle Z(k,t) Z(-k,t) \rangle$, where $Z(k,t)$ is the Fourier transform of $z(x,t)$, and the q-order height-height correlation function,[7]

$$\sigma_q(\Delta,t) = \left\langle \overline{|z(x,t) - z(x+\Delta,t)|^q} \right\rangle_R^{1/q}.$$

When initially a flat interface starts to roughen the global width of interface shows the following behavior [1]. At the initial stage, $W(L,t)$ grows as a power law of time, $W(L,t) \propto t^\beta$, where $\beta$ is called the growth exponent, while at the later stage, $W(L,t)$ saturates to a certain power of the system size $L$, $W(L,t) \propto L^\alpha$, where $\alpha$ is the global roughness exponent. The crossover to the saturated state is governed by the lateral correlation length $\xi(t)$, which scales as $\xi(t) \propto t^{1/z}$ at the initial stage $t \ll L^z$ and saturates to $L$, when $t \gg L^z$, where $z = \alpha/\beta$ is the so-called dynamic exponent. This behavior is known as the Family- Vicsek dynamic scaling *ansatz* [1].

In the absence of any characteristic length scale in the system, the interface roughening displays scaling invariance, *i.e.*, it does not change under rescaling of space and time combined with an appropriate rescaling of the observables and the control parameters [1]. In this case

$$w(\Delta,t) \propto t^\beta f_w(\Delta/\xi(t)), \ \sigma_q(\Delta,t) \propto t^\beta f_q(\Delta/\xi(t)), \qquad (1)$$

and

$$S(k,t) \propto k^{-(2\alpha+1)} f_S(k\xi(t)), \qquad (2)$$

where the scaling functions behave as $f_i(y) \propto y^\alpha$ ($i = w, q$) and $f_S(y) \propto y^{2\alpha+1}$, if $y \ll 1$, and they become constants when $y \gg 1$ [1-6]. So, the saturated self-affine roughness is characterized by the unique scaling exponent $\alpha = H \leq 1$, also called the Hurst exponent, *i.e.*,

$$W(L) \propto L^H, \quad \sigma_q(\Delta) \propto w(\Delta) \propto \Delta^H, \tag{3}$$

and

$$S(k) \propto k^{-(2H+1)}. \tag{4}$$

The self-affine interfaces were found in a large variety of dynamic systems [1-6]. Generally, however, in a system far from equilibrium, the system size itself can play the role of control parameter that governs the spatiotemporal dynamics of the system.[8][9][10] Accordingly, the interface roughening dynamics may also depend on the system size $L$.[11]

A simple example of this phenomenon is the so-called anomalous kinetic roughening, characterized by different scaling exponents in the local and the global scales [11][12][13][14][15]. In such a case, the local roughness amplitude depends on the system size. Moreover, the power spectrum of growing interface also may be size dependent; nevertheless, the global width displays the Family-Vicsek dynamic scaling *ansatz*, *i.e.*, the behavior of different scaling functions are generally governed by different scaling exponents.

According to the concept of the generic dynamic scaling in kinetic roughening [13], the scaling functions behave as

$$f_w \propto y^\zeta \text{ and } f_q \propto y^{\zeta_q}, \text{ if } y \ll 1, \text{ while } f_w = const \text{ and } f_q = f(q), \text{ when } y \gg 1, \tag{5}$$

whereas

$$f_S(y) \propto y^{(2\alpha_S+1)}, \text{ if } y \ll 1, \text{ and } f_S(y) \propto y^{(\alpha-\alpha_S)}, \text{ if } y \gg 1, \tag{6}$$

where $\zeta < \alpha$ is the local roughness exponent, $\zeta_q$ is the multi-affine spectrum of local roughness exponents ($\zeta_2 = \zeta$) [8], and $\alpha_S$ is the spectral roughness exponent [13].

Therefore, the local characteristics of saturated anomalous roughness depend on the system size as [12-15]

$$w(\Delta, L) \propto \Delta^{\zeta} L^{\alpha-\zeta}, \quad \sigma_q(\Delta, L) \propto \Delta^{\zeta_q} L^{\alpha-\zeta_q}, \tag{7}$$

and

$$S(k, L) \propto k^{-(2\zeta_S+1)} L^{\theta}, \tag{8}$$

where

$$\theta = 2(\alpha - \zeta_S). \tag{9}$$

Accordingly, for self-affine interfaces $\zeta = \zeta_S = \alpha$, while the intrinsically anomalous roughness is characterized by $\zeta = \zeta_S < \alpha$; the super-roughness is characterized by $\zeta = 1, \zeta_S = \alpha > 1$, while the faceted roughness is also characterized by $\zeta = 1$, but $\alpha > \zeta_S > 1$ [13]. Multi-affine roughness is characterized by the spectrum of scaling exponents $\zeta_q$, such that $\zeta_2 = \alpha$, whereas in the case of anomalous multi-scaling $\zeta_q \leq \alpha = const$ for all $q$ [15].

Commonly it is assumed that the scaling exponents are constants, *i.e.*, they do not depend on the variables ($x,t,L$) [1-15]. However, the roughness of some types of interfaces is characterized by the scaling exponents which are not constants; rather they are continuous functions of control parameters.[16][17] Namely, more generally, the local width behaves as

$$w(L,\Delta,t) \propto \Phi_w(\Delta^{\zeta(L,\Delta,t)}, t^{\beta(L,\Delta,t)}, L^{\alpha-\zeta(L,\Delta,t)}), \qquad (10)$$

instead of simple power-law scaling. In this case, to produce a data collapse, the functional dependences of scaling exponents should satisfy a group homomorphism, which can be expressed as a set of partial differential equations associated with the concept of local scaling invariance [16].

Dynamic scaling with continuously varying exponents was observed in certain experiments in turbulence[18] and in some DLA-related growth processes.[19] In both cases the roughness exponent and growth exponents depend on time and the window size $\Delta$. The aim of this work is to show that in systems far from equilibrium the local roughness exponent may be system size dependent.

## 2. EXPERIMENTS

We are interested in the system size dependence of the roughness exponent. Accordingly, we note that in experiments with turbulent premixed propane/air flames were observed the increase in the fractal dimension of flame fronts as the Reynolds number, $Re = Lv/\mu$, increases [20] (here $v$ is the mass velocity and $\mu$ is the kinematic viscosity). So, a promising candidate to study a system size dependent roughening is a turbulent flame front formed in paper combustion experiments.[21]

A definitive advantage of experiments with paper is that these experiments are easy to perform and the interface configurations formed in paper can be well defined. Recently, papers were widely used to study the kinetic roughening of interfaces formed in paper wetting [2,22], fracturing [2,22,23], and burning experiments [21,24,25,26,27,28,29]

### 2.1 Experimental details

Early, the kinetic roughening in slow combustion of paper was studied in works [24-29] in which the flame dynamics was essentially laminar.[30] We note that the forced air flow is essential to produce the flame turbulence [21]. So, to obtain a turbulent flame, in this work the sheets of paper were burned in a combustion chamber with forced air flow along a horizontally oriented burned sheet.

In this work we were especially interested in the effect of sheet width on the turbulent flame front roughness, rather than in the flame roughening dynamics. Accordingly, to simplify the image proceeding and reduce the uncertainty in the flame front definition, we studied the post-mortem marks of flame, formed when the burning is quenched (see Fig. 1). To obtain well defined flame front marks we used the relative thick papers the burning of which is accompanied by a high smoke. So we were not able to film the burning process *in situ* with a satisfactory image quality.

To keep the paper sheet planar during combustion, it was placed in the open metallic frame in such a way that one end of the sheet is fixed in the holder and the lateral edges of burning zone are kept by thin bridges attached to the paper edges fixed in the frame (see Fig. 2). The papers were ignited from free edge using an electrical heating wire stretched across the paper sample. Air flow in the direction of paper burning was produced by a by a cross-flow blower. The combustion was quenched when the burning front achieves the mark on the middle of sheet (see Fig. 2).

In order to study the effects of the paper properties on the flame front, in this work the combustion experiments were done on sample sheets of two different papers, early used in Ref. [22] to study the scaling properties of rupture lines and wetting fronts:[31] the "Secant" paper (sheet thickness is 0.338±0.04 mm; basis weight of 200±30 g/m$^2$) and the "Filtro" paper with open porosity (sheet thickness 0.321±0.05 mm; basis weight of 130±20 g/m$^2$). Both papers are characterized by the well pronounced anisotropy associated with preferred fiber alignment along the paper making machine direction (see

Ref. [22]). In this work we studied the interfaces with middle line across the machine direction of paper.

Examples of interfaces formed in these papers in the imbibition, fracture, and burning experiments are shown in Figs. 3 and 4. The scaling properties of saturated wetting fronts and rupture lines were reported in [22]. In this work we studied the saturated roughness of the post-mortem flame and burning fronts in sheets of different width $L$ (see Fig. 5), which was varied from $L_0 = 10$ mm to $L_{max} = 50$ cm, with the relation $L = \lambda L_0$ for scaling factors $\lambda =$ 2.5, 5, 7.5, 10, 15, 20, 30, 40, and 50, whereas the length of all paper sheets used was 40 cm. At least 30 burning experiments were with sheets of each size. So the data reported below are averaged over these experiments.

We found that the flame is laminar, if the forced air flow velocity $v_a$ is less than the velocity of burning front $v_b$ and it becomes turbulent when $v_a > v_b$. The burning velocity increases slightly when the flame becomes turbulent. In our experiments the burning velocity without forced air flow was $v_b \approx 2.0 \pm 1.0$ mm/s and it achieves $v_b \approx 2.5 \pm 1.5$ mm/s, when $v_a = 5$ mm/s.[32] No systematic change in the combustion velocity as the sheet width increases was found.

We also note that the behavior of downward flame spread over a paper sheet depended not only on the air stream velocity but also on its changing rate (see also[33]) in such a way that the burning may be quenched when the air flow velocity $v_a > 4v_b$. However, the effect of air flow rate on the fire front dynamics was not studied in this work, because of the rising of hotter air (convection) due to paper combustion also affects the flame dynamics and makes difficult to control the local air flow velocity. Accordingly, we control only the mean velocity of air flow produces by the air blower which was around of $v_a = 5 \pm 2$ mm/s. $> v_b$ in all experiments reported below.

**2.2 Image proceeding**

For quantitative studies all burned sheets were scanned with 700x700 ppp resolution (see Figs. 1, 4 - 5) and then each interface was digitized (see Figs. 4 b, c and 5) using Scion Image software [34] and plotted as a single-valued function [35] $z_L(x)$, in xls-format (see Figs. 6 and 7). The pixel size is 0.04 mm, that is close to the mean fiber width (see [22]) and it is well bellow of the average length of fibers, which is $\ell_F \approx 3$ mm.

It should be emphasized that in this work we were interested in the small-scale local roughness ($\Delta \leq 6$ mm $\cong 2\ell_F$) associated with the flame turbulence, in contrast to other works [24-29], which were focused on the roughness at the scales larger than fiber length.[36]

**2.3 Scaling analysis**

The configurations of burning and flame fronts are assumed to be described by the single-valued functions [35] $z_i(x)$ (see Figs. 6, 7). To eliminate the effect of sheet borders on the front configurations, the scaling analysis was performed only on the central parts of interfaces, leaving the edges of width $0.2L$ (see Fig. 6).

The scaling properties of each interface were analyzed using three techniques adopted in the BENOIT 1.3 software:[37] the rescaled-range analysis, $R/S \propto \Delta^{\zeta}$, the variogram, $V \propto \Delta^{2\zeta}$, and the power spectrum, $S \propto k^{-(2\zeta_s+1)} L^{2\theta}$, methods.[38] The global roughness exponents were determined from the scaling behavior of interface width (3).

Furthermore, to check and quantify the multiscaling properties of the burning and fire fronts, we have studied the scaling behavior (7) of q-order height-height correlation function for $q = 1,2,...,10$ (see Ref. [7]).

**2.4 Empirical results**

First of all we found that the burning fronts in both papers display the multi-affine scaling behavior which is characterized by the same spectrum of local roughness exponent (see Figs. 8, 9, 10 and 11 a):

$$\zeta_q = 0.93 q^{-0.15} \text{ for } q \geq 1 \text{ and } \alpha = \zeta_2 = 0.83 \pm 0.04; \quad (11)$$

and the structure factors of burning fronts are found to be size independent ($\alpha_S = \alpha = \zeta_2$ and $\theta = 0$).

We note that the value of $\zeta_2 = 0.83 \pm 0.04$ is consistent with the small-scale ($\Delta < 2$ mm) roughness exponent $\zeta = 0.88 \pm 0.05$ of moving combustion front which was reported in Ref. [26], where it has been found that at larger scales the combustion front roughness follows the scaling predictions of the Kardar-Parisi-Zhang equation with thermal noise, i.e., $\zeta(5 < \Delta < 100) = 0.5$. The power-law decrease of $\zeta_q$ as $q$ increases is also consistent with data for small-scale roughness reported in Ref. [27].

On the other hand, we found that the flame front roughness displays an anomalous multi-scaling, characterized by the global roughness exponent and the sheet width dependent spectra of local roughness exponents, $\zeta_q(\lambda)$ (see Figs. 11 b and 12, 13). Specifically, we found that

$$\zeta_q(\lambda) = \zeta_1(1) q^{-\omega} \lambda^{-\phi} < \alpha, \ \zeta(\lambda) = \zeta_2 = \zeta_2(1)\lambda^{-\phi}, \tag{12}$$

where $\zeta_1(1), \zeta_2(1)$ are constants and $\omega$ and $\phi$ are new scaling exponents (see Table 1).

We note that the roughness of flame fronts in different papers is characterized by different global roughness exponents and different spectra of local roughness exponents $\zeta_q(\lambda)$ (see Table 1), nevertheless the spectral roughness exponent is found to be

$$\alpha_S = 0.50 \pm 0.05 < \alpha, \tag{13}$$

for the flame fronts in both papers.

We also found that the power spectrum of flame fronts behaves as (see Fig. 14):

$$S \propto k^{-2}L^{2\theta}, \quad (14)$$

where (see Table 1)

$$2\theta \cong -\phi < 0 < 2(\alpha - \zeta_S); \quad (15)$$

*i.e.*, $S(k,L)$ decreases as the sheet width increases; the symbol $\cong$ means the numerical (experimental) equality. This behavior differs from that expected in the case of anomalous kinetic roughening (see Eqs. (8-9) and Ref. [13 15]).

Furthermore, we found that the correlation length of the q-order height-height correlations also possesses unconventional scaling behavior (see Figs. 15, 16):

$$\xi_q(\lambda) = \xi_1(1)q^{-\mu}\lambda^{\nu} \text{ for } q \geq 1, \quad (16)$$

where, numerically

$$\mu \cong \omega \text{ and } \nu \cong \alpha; \quad (17)$$

see Table 1.

## 3. DISCUSSION

Paper combustion is a complex process involving many chemical reactions along with complicated air flows. In this work we are interested only in the system size dependence of turbulent flame front roughness which is closely related to the intensity of gas-flow turbulence and the nonuniformity of combustible burning media.[39] Accordingly, the turbulent flame can be represented by means of the Navier-Stokes set of differential equations or their extended forms.[40]

Turbulence is essentially multiscale phenomenon associated with the Kolmogorov cascade of energy flux from the largest scale to the smallest one.[41] This cascade is commonly modeled by multiplicative processes, generically leading to multifractal fields [[42]]. As a result, the flame front obeys a multi-affine geometry.

A continuous-scale limit of multiplicative cascades leads to the family of log-infinitely divisible distributions, among which are the universal multifractals which have a normal or Lévy generator, and for which [43]

$$\zeta_q = \zeta_1 - \frac{C}{\rho - 1}\left(q^{\rho-1} - 1\right), \qquad (18)$$

where $C \leq d$ is an intermittency parameter and $0 < \rho \leq 2$ is the basic parameter which characterizes the process (Lévy index; $\rho = 2$ corresponds to the log-normal distribution).

From comparison of equations (11) and (18) follows that the burning fronts in both papers are related to multiplicative cascade with $\rho = 0.85$ and $C = 0.15\zeta_1 = 0.14 \pm 0.03$.

On the other hand, from comparison of equations (12) and (18) follows that the turbulent flame may be associated with the Lévy-stable process such that

$$\rho = 1 - \omega \qquad (19)$$

and

$$C(\lambda) = \omega \zeta_1(\lambda) = C(1)\lambda^{-\phi} \qquad (20)$$

are dependent on the properties of burning media. Specifically for flames in the "Secant" paper $\rho = 0.8 \pm 0.04$, $C(1) = 0.14 \pm 0.04$, while for flames in the "Filtro" paper $\rho = 0.57 \pm 0.05$ and $C(1) = 0.37 \pm 0.07$ (see Table 1).

To take an insight into the physics of the observed dependence of local roughness exponent on the system size, we note that when the kinetic energy fluctuations $\Omega$ in a system far from equilibrium exceed a certain critical value $\Omega_C$, the system dynamics is characterized by system size dependent largest Lyapunov exponent,[44]

$$\Lambda(L) \propto P^\varphi \propto L^{-\gamma}, \qquad (21)$$

where $P$ denotes the perturbation (the power consumption of turbulent flow,[45] in the present case), and $\varphi, \gamma > 0$ are the scaling exponents determined by the microscopic nature of the system. This implies that the fractal dimension of strange attractor,[46] $D_A \propto \Lambda$, as well as the fractal dimension of avalanches,[47] $D_{Av} \propto D_A$, decrease as the system size increases. Accordingly, the local roughness exponent of turbulent front, $\zeta_2 = D_{Av} - d$ [46], also should be a function of system size, $e.g.$,

$$\zeta_2 \propto L^{-\gamma}. \tag{22}$$

From the comparison of Eq. (22) and the second (empirical) relation in Eq. (12) follows that

$$\phi = \gamma. \tag{23}$$

Now one may speculate that the intermittency parameter $C(\lambda)$ in Eq. (18) is related to the largest Lyapunov exponent, $e.g.$, $C(\lambda) \propto \Lambda(\lambda)$.

The unusual dependence of the structure factor leaves to future work the important questions concerning the numerical equalities (15) and (17).

## 4. CONCLUSIONS

The kinetic roughening of interfaces in systems far from equilibrium depends on the system size. Specifically, we show that the roughness of turbulent interfaces may be characterized by system size dependent spectrum of local roughness exponents. This gives a rise to a new type of scaling behavior related to the Lévi-stable multiplicative process with system size dependent intermittency parameter. We expect that the system size dependence of scaling exponents should be observed in a wide variety of systems far from equilibrium, when $\Omega > \Omega_C$.[48]

**Acknowledgments**

This work was supported by the Mexican Government under the CONACyT Grant No. 44722 and by the Mexican Petroleum Institute under Chicontepec research program. The authors thank Miguel A. Rodríguz and Juan M. López for useful discussions.

[48] For example, the authors of [*] have observed that fractal dimension of atmospheric aerosol aggregates increases from 1 to above 2 as the aggregate size increases.

**Table 1. Scaling exponent for turbulent flame fronts in different papers.**

| Scaling exponent | "Filtro" paper | "Secant" paper |
| --- | --- | --- |
| $\alpha$ | 1.00±0.04 | 0.83±0.03 |
| $\zeta_1(1)$ | 0.86±0.06 | 0.72±0.05 |
| $\zeta_2(1)$ | 0.77±0.04 | 0.63±0.03 |
| $\omega$ | 0.43±0.05 | 0.20±0.03 |
| $\phi$ | 0.10±0.03 | 0.38±0.05 |
| $2\theta$ | -0.12±0.05 | -0.40±0.06 |
| $\mu$ | 0.43±0.05 | 0.20±0.02 |
| $\nu$ | 1.00±0.08 | 0.83±0.08 |

**Figure 1.** Scanned images of the burned sheets ($L = 50$ mm) of "Secant" paper after: a) slow combustion without forced air flow; b) slow combustion with forced air flow velocity $v_a = 5\pm2$ mm/s. $> v_b$.

**Figure 2.** Configuration of paper samples used for slow combustion experiments.

**Figure 3.** Scanned images of rough interfaces in the sheets of "Secant" paper of width $L = 100$ mm, obtained in: a) slow combustion; b) tensile fracture [ ]; and c) imbibition [ ] experiments.

**Figure 4.** Interfaces formed in the "Filtro" paper sheets of width $L = 100$ mm: a) grey-scale scanned images, b) digitized black-and-white images, and c) contours.

**Figure 5.** Postmortem burning and flame fronts in the "Filtro" paper sheets of different width: a) $L = 300$ mm (the snapshot shows the scanned image and the graphs of postmortem fronts in the sheet of width $L = 10$ mm); b) $L = 150$ mm; c) $L = 75$ mm. d) Digitized graphs of the burning and the flame fronts shown in Fig. c.

**Figure 6.** Graphs of the burning (1) and flame (2) fronts, the rupture lines (3) and the wetting front (4) in "Filtro" paper (see Figure 4).

**Figure 7.** Graphs of burning (1) and flame (2) fronts in the sheets of "Secant" paper of width: a) $L = 100$ mm and b) $L = 500$ mm.

**Figure 8.** a) Log-log graphs of a) $R/S$ and b) $V$ versus $\Delta$ for the quenched burning fronts in "Secant" (1,2) and "Filtro" (3,4) paper sheets of width $L = 25$ mm (1,3) and $L = 500$ mm (2,4). The graphs are shifted for clarity.

**Figure 9.** Power spectrum of burning fronts in "Secant" (1,2) and "Filtro" (3,4) paper sheets of width $L = 10$ mm (1,3) and $L = 500$ mm (2,4). The graphs are shifted for clarity.

**Figure 10.** a) Log-log graphs of the q-order height-height correlation function of the postmortem burning fronts in "Filtro" paper (from bottom to top: $q = 1$, 2, 3, 6 10); b) spectrum of local roughness exponents of burning fronts in "Secant" (circles) and "Filtro" (triangles) papers.

**Figure 11.** Log-log graphs of the global front width versus sheet width for: a) burning and b) flame fronts in the "Secant" (1) and "Filtro" (2) papers.

**Figure 12.** Variogram graphs for the postmortem flame fronts in a) "Secant" and b) "Filtro" paper sheets of width $L = 10$ mm (1), $L = 100$ mm (2), and $L = 500$ mm (3). The graphs are shifted for clarity.

**Figure 13.** a) The local roughness exponent $\zeta_2$ versus $\lambda$; and b) the sheet width dependence of correlation length $\xi_2$ for the flame fronts in "Secant" (1) and "Filtro" (2) papers.

**Figure 14.** a) Power spectrum of flame fronts in the sheets of "Secant" (1,2) and "Filtro" (3,4) papers of width $L = 500$ mm (1,3) and $L = 10$ mm (2,4). b) System size dependence of $A = Sk^2$ for the flame fronts in "Secant" (1) and "Filtro" (2) papers.

**Figure 15.** Log-log graphs of the q-order height-height correlation function of the postmortem flame fronts in a) "Secant" and b) "Filtro" paper sheets of width $L = 100$ mm (from bottom to top: $q = 1$, 2, 3, 6 10). The graphs are shifted for clarity.

**Figure 16.** a) Data collapse for the q-spectra of local roughness exponents of the flame fronts in "Secant" (1) and "Filtro" (2) papers ($\Psi_q = \zeta_q \lambda^\omega$). b) Log-log graphs of $\xi_q^* = \xi_q \lambda^{-\alpha}$ versus $q$ for the flame fronts in "Secant" (1) and "Filtro" (2) papers.

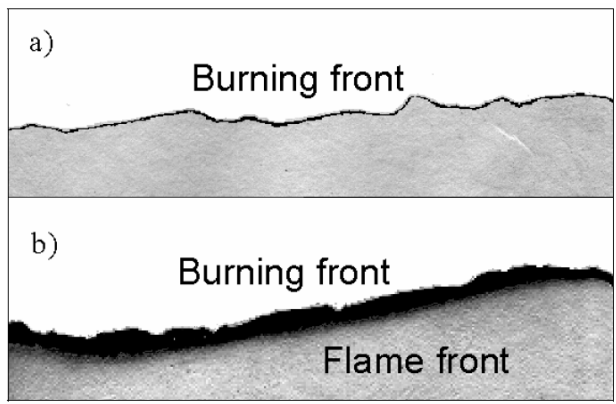

Figure 1.

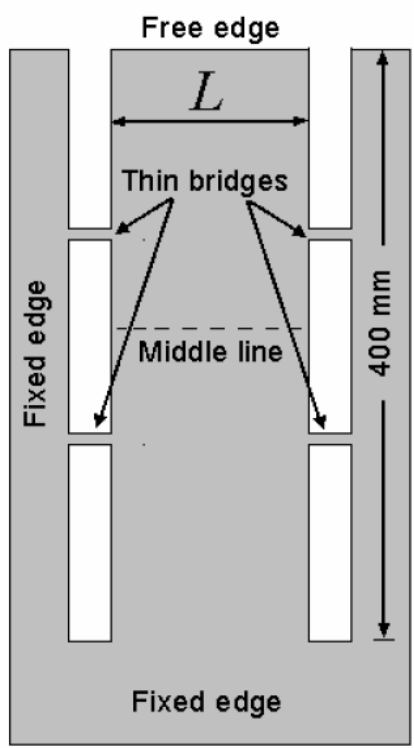

Figure 2.

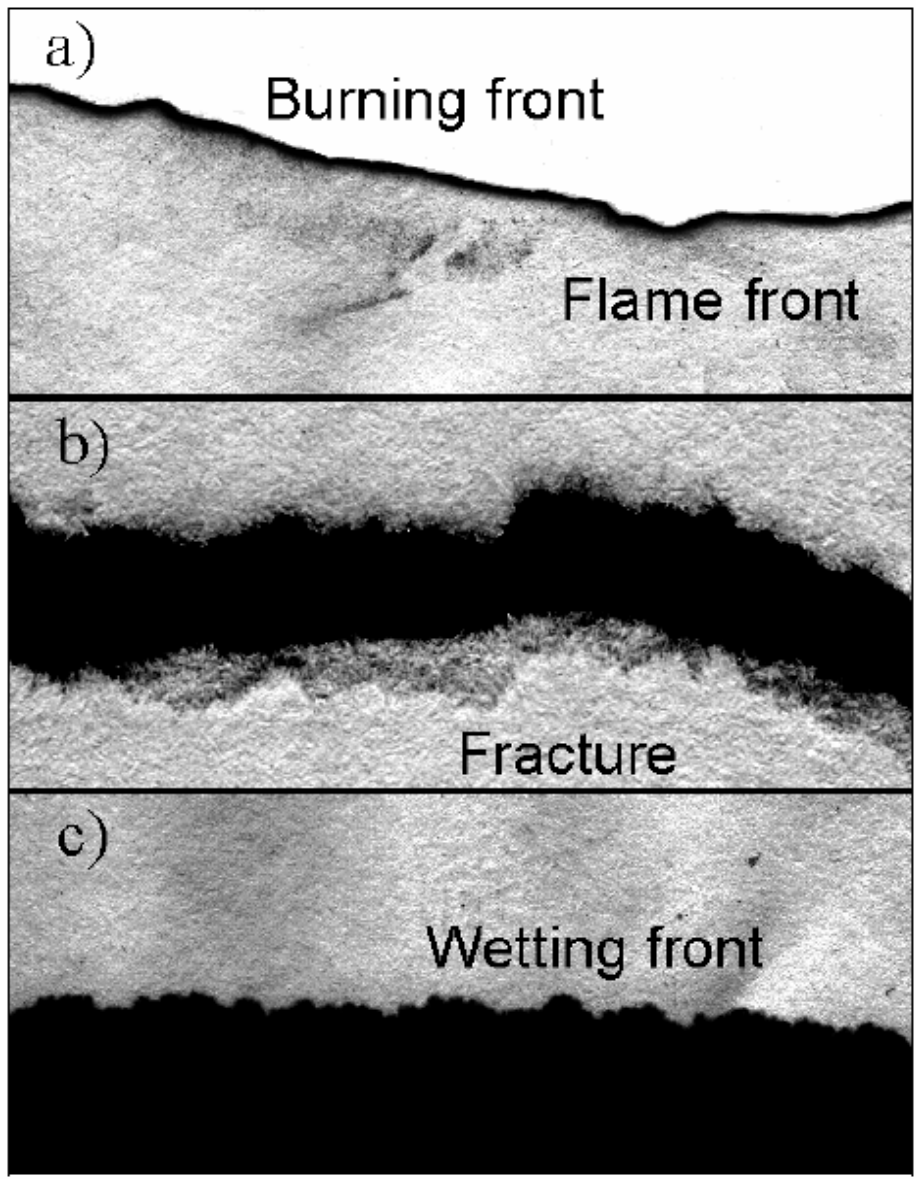

Figure 3.

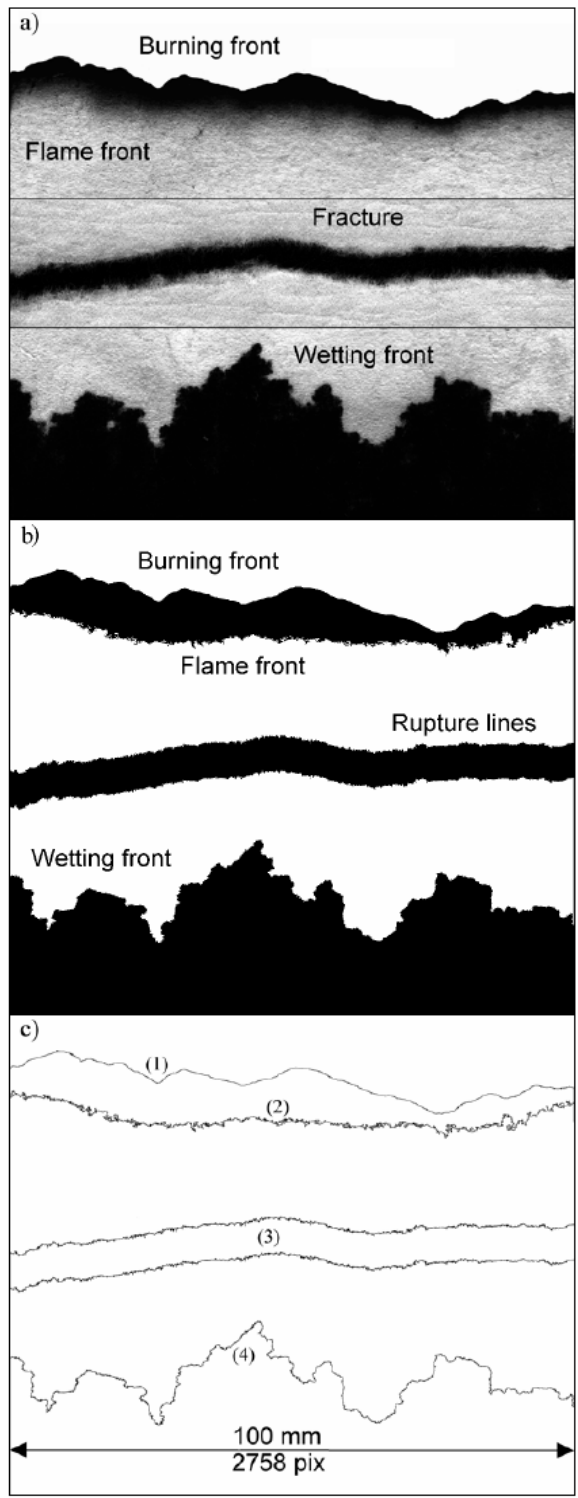

Figure 4.

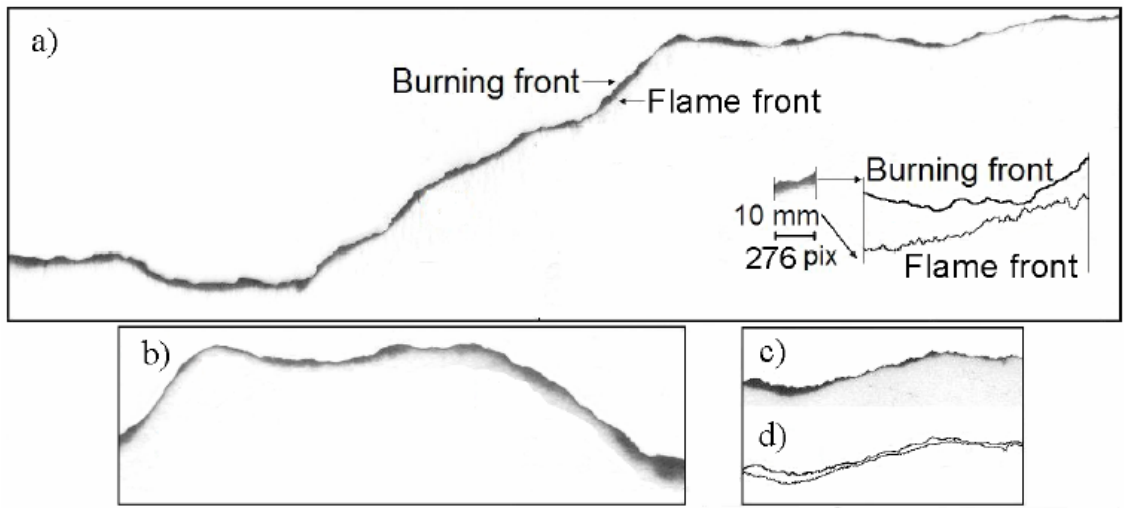

Figure 5.

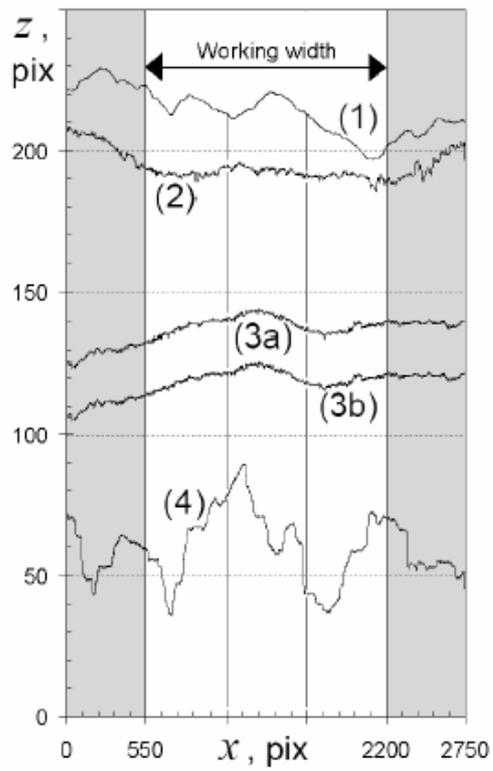

Figure 6.

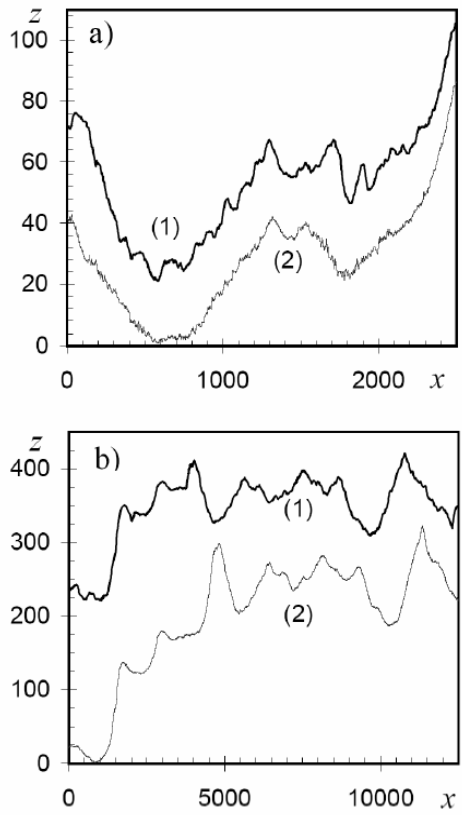

Figure 7.

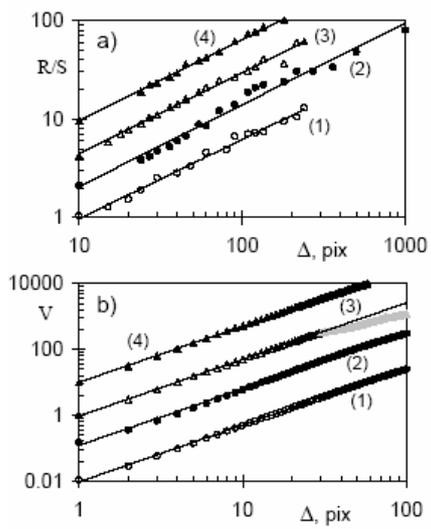

Figure 8.

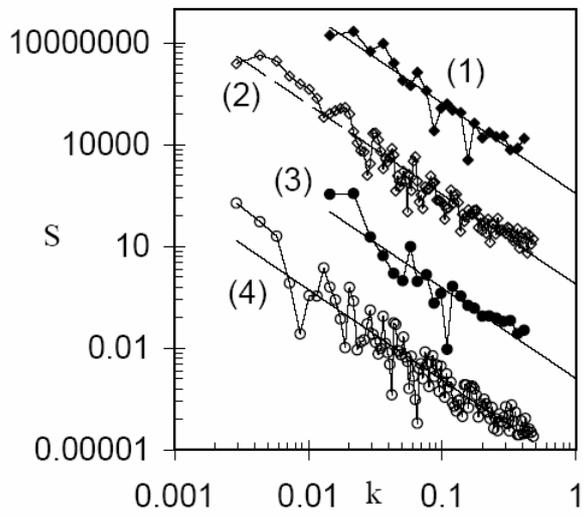

Figure 9.

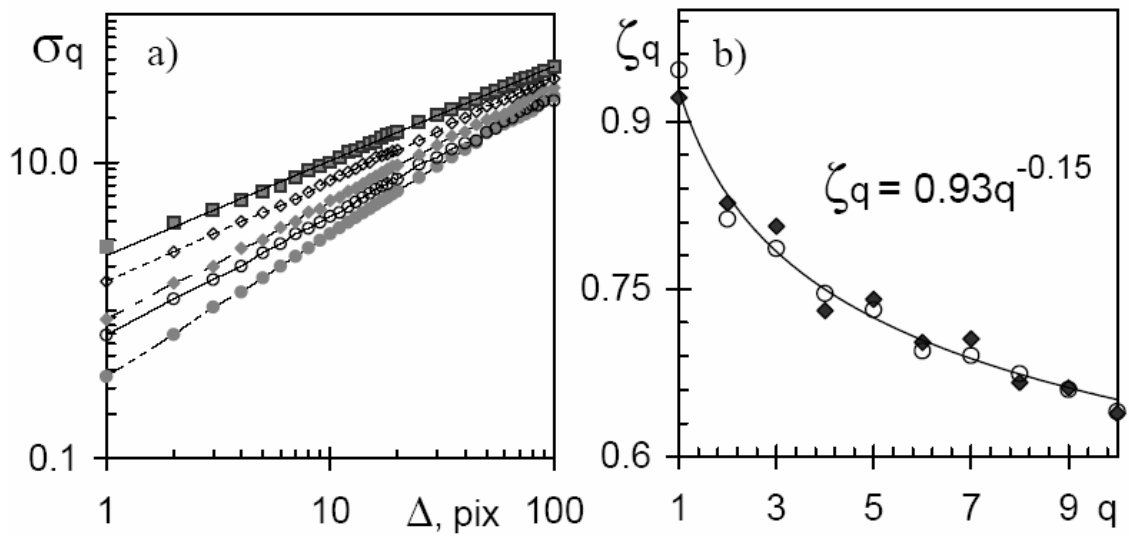

Figure 10.

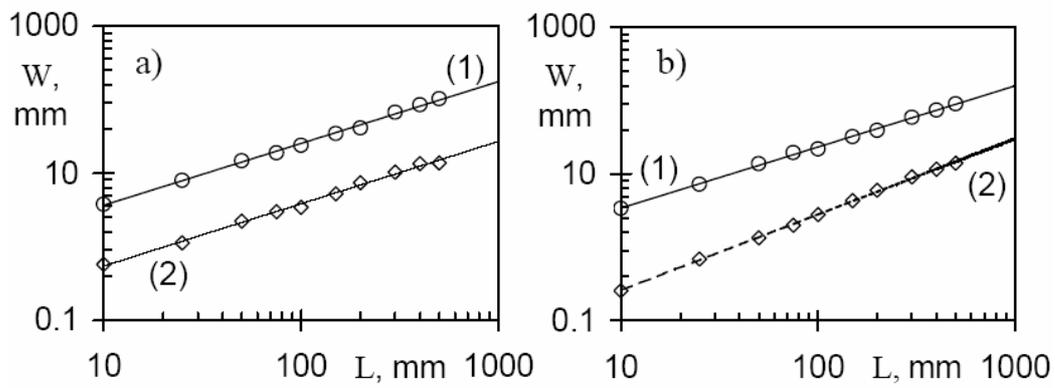

Figure 11.

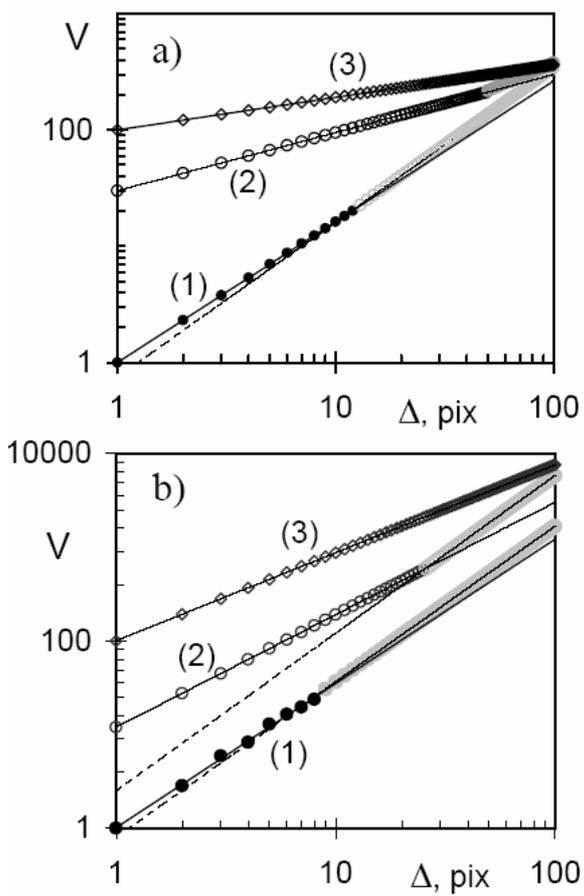

Figure 12.

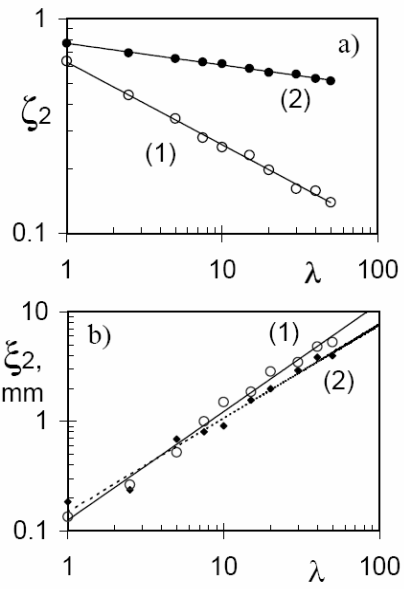

Figure 13.

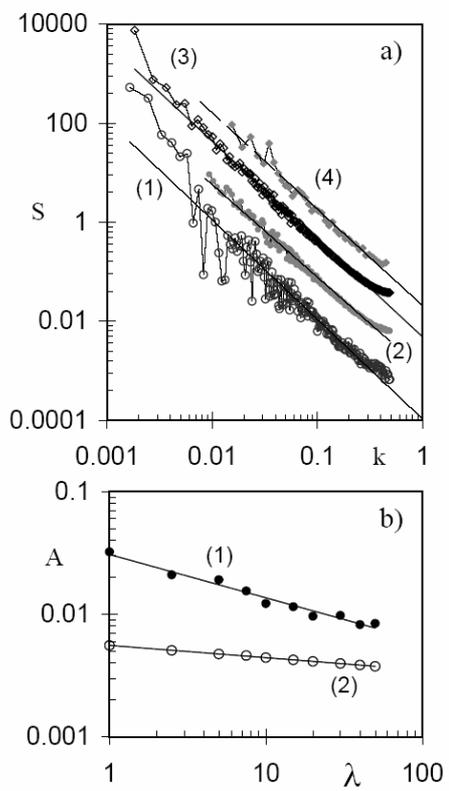

Figure 14.

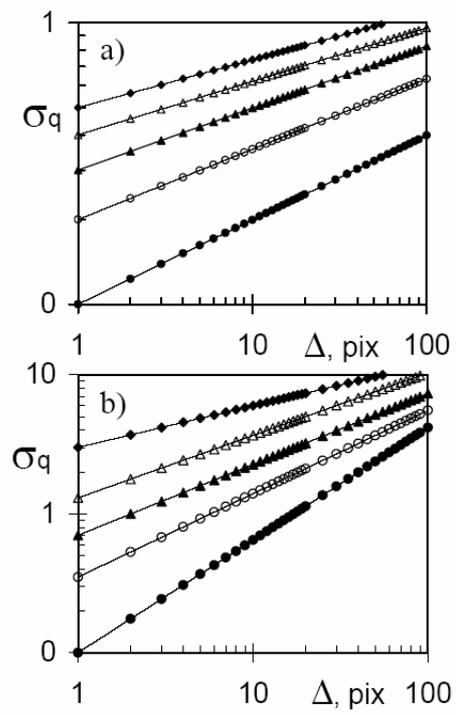

Figure 15.

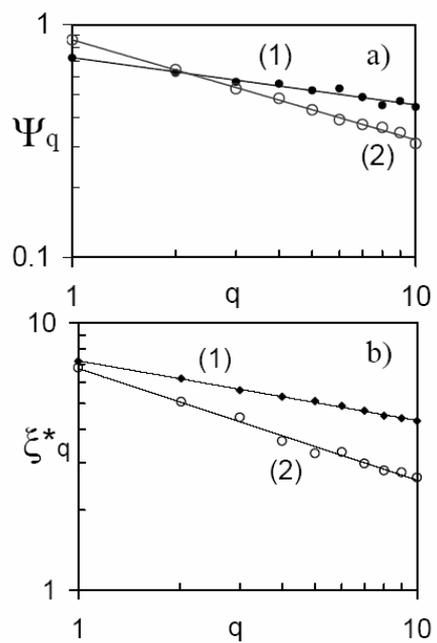

Figure 16.